# A wide-angle metamaterial narrow-band-stop filter for 532 nm wavelength green light

Liyang Yue[1,*], Songkun Ji[1], Bing Yan[1], Nguyen Thanh Tung[2], Vu Dinh Lam[2] and Zengbo Wang[1,*]

[1]*School of Electronic Engineering, Bangor University, Dean Street, Bangor, Gwynedd, UK, LL57 1UT*
[2]*Institute of Material Science, Vietnam Academy of Sciences and Technology, 18 Hoang Quoc Viet, Cau Giay, Hanoi, Vietnam*
[*]*l.yue@bangor.ac.uk* and *z.wang@bangor.ac.uk*

**Abstract**
Traditional optical interference narrow-band-stop filters do not possess wide-angle property, because peaks and troughs of filter spectrum would be moved at a non-normal angle of incidence (AOI), which could result in functional failure in particular cases, e.g. blocking of laser for pilot in cockpit during premeditated laser pointer direct. For this reason, we designed a wide-angle metamaterial narrow-band-stop filter assembled by cross shaped units to block 532 nm green light, which is firstly reported in the world. Unnecessary shift of spectrum caused by AOI change is effectively inhibited, and angular tolerance of wide-angle capability achieves to ±35 ° non-normal AOIs.

## 1. Introduction

Interference narrow band filters structured as multiple thin coatings or films on glass are commonly used to achieve bandpass/stop functions in visible wavelength range at normal 0 ° angle of incidence (AOI), while change of AOI is considered as a method to shorten central wavelength of the interference filter and achieve partial tenability for filter, because of shift and widening of transmission band [1]. However, similar change of filter band makes against device's wide-angle performance which is first priority for certain applications, e.g. development of an airplane windscreen to block laser pointer irradiation. It requests that main features of filter spectra, e.g. troughs or peaks, have not to be moved to other wavelengths to ensure same filtering effect at different AOIs. At this point angular tolerance of traditional bandpass/stop filter is less than 1 °, and there has not been reported about a wide-angle band filter for visible spectrum so far. Meanwhile, metamaterial is able to gain electromagnetic characteristics have not been found in nature from its structure arranged by repeating patterns rather than base material property [2]. Metamaterial band-stop filter is a type of metamaterial to efficiently block electromagnetic radiation at certain frequency band with wider adaptability and better effectiveness, and split ring resonator (SRR) [3, 4] and following complementary split ring resonator (CSRR) [5] were earliest works and most frequently applied structures to make it. Subsequently various modified designs based on them were emerging, and their resonant types were investigated in this process [6-9]. One of them, three dimension (3D) 'Fishnet' metamaterial, firstly studied by Valentine et al. in 2008 [10] effectively simplifies the shape of metamaterial cell and reduce difficulties of fabrication. Moreover, 'Fishnet' shape can be thought as many tiny crosses cascaded, and metamaterial made of cross shaped units was found to have band pass/stop capability. Paul et al. invented a cross shaped metamaterial band-pass filter operating in THz band and pronounced over 80% peak amplitude transmission in the passband in 2009 [11]. For near optical frequency a band-stop filter structured as a coupled cut-wire was designed by Li et al. [12], however no wide-angle property claimed in this study. Therefore, compared to GHz, THz metamaterial filter, development of a wide-angle metamaterial narrow-band-stop filter functions in visible spectrum is more challenging because cell structure needs to sustain of similar resonance for shorter wavelengths at non-normal AOIs.

In this paper, we successfully design a wide-angle metamaterial narrow-band-stop filter made of silver (Ag) to block 532 nm wavelength green light using finite integration technique (FIT) software (CST studio). Setting of 532 nm wavelength as the narrow-stop-band is due to existing laser hazard in aviation environment, because most of powerful diode pumped solid state (DPSS) laser pointers threatening of pilot are 532 nm wavelength green laser nowadays. Based on this practical demand, a numerical transmission model is built to evaluate proposed filter's capability at multiple non-normal AOIs. Unit cells are regularly aligned Ag crosses with same spacing, which is inspired from a metamaterial THz perfect absorber reported by Liu et al. [13], however cross size is reduced to adapt much smaller operating wavelength. Computing wavelength, $\lambda$, for this study is from 400 to 1100 nm, and mechanism of wide-angle capability is explained by near-field power flow analysis.

## 2. Modelling

Geometrically this novel metamaterial filter is structured as a quartz glass covered by regularly arranged nanoscale Ag crosses. A model of single cross cell is created by CST studio using tetrahedral meshing for frequency domain solver, and regarding optical properties, e.g. refractive index, $n$ and extinction coefficient, $k$, for Ag and quartz glass are from the data in previous publications [14, 15]. The boundary conditions of model is selected as 'unit cell' in $x$ and $y$ axis directions, which means the computing boundaries at $x$ and $y$ direction will be simulated as that contact to the identical structure with current model and extend this repeatable pattern to the infinite space to mimic metamaterial cells on glass substrate, as shown in figure 1. Two ports, $Z_{max}$ for incidence and $Z_{min}$ for acceptance, are automatically generated at upper and lower sides of model to collect scattering parameters for plotting of transmission curve. Their directions of electric and magnetic field, $v$ and $u$, are along $y$ and $x$ axes respectively as indicated in inset of figure 1. Meanwhile, boundary condition of $z$ axis direction is configured to 'open' and background medium in the space is set to air for approaching of reality. Cross geometry are defined by four parameters, height, $h$, width, $w$, length, $l$ and spacing between each other, $a$, also AOI is defined as an angle called *theta* in the boundary condition setting, as shown in inset of figure 1. Besides another model representing common quartz glass without metamaterial structures on the surface is built under the same conditions, which demonstrates transmissions of glass substrate and provides a reference for comparisons. Transmissions of current model are varied by multiple AOIs to exhibit outstanding wide-angle capability within calculation range ($\lambda$, 400 - 1100 nm).

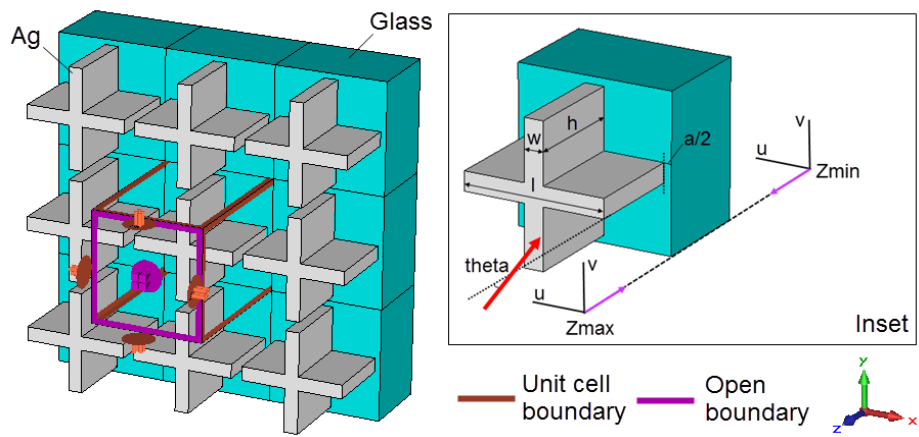

Figure 1 CST model and applied boundary conditions with inset of geometrical parameters and scattering parameter ports

## 3. Result
### 3.1 Pure glass and normal AOI

Figure 2 (a) shows that band filter spectra of proposed metamaterial structure with optimized geometric parameters (a = 28 nm, h = 198 nm, l = 264 nm and w = 36 nm) at multiple AOIs. Meanwhile, transmission curve of quartz glass without metamaterial feature as a reference (pink dash line) is also included in figure 2 (a). Average over 0.85 transmission is observed at reference curve, even nearly 1.0 transmission within 500 - 600 nm wavelength green light range. On the contrary, transmission (band filter spectra) for metamaterial filter covering on same glass transmission is lower than this level at normal AOI (theta = 0 °, black line in figure 2 (a)), nevertheless a most obvious sharp trough at 532 nm wavelength (black dash line window) emerges on theta = 0 ° curve in the wavelength section where highest transmission of reference glass model should locate in, which means metamaterial structure is capable to change property of glass and allow original green light to be filtered to low transmission. For normal AOI (theta 0 °) the bottom of the trough arrives to 0.13 transmission at 532 nm wavelength (black line in the inset of figure 2 (a)), then transmission will increase by 90% around ±5 nm wavelength from the bottom (532 nm). The declining and following rebound of trough for blocking of 532 nm wavelength almost concentrate in green light wavelength range (500 - 570 nm), and basic 0.60 transmission can be secured for neighbouring blue and yellow light on the visible light spectrum. After 600 nm wavelength transmission curve of theta 0 ° will go down and then form a small peak at 970 nm wavelength in near infrared (IR) range.

### 3.2 Non-normal AOIs

Proposed metamaterial filter enables to maintain its filter spectra at multiple AOIs. Multiple transmission curves representing non-normal AOIs are illustrated in figure 2 (a) and its inset on the right. Curves of theta 15°, 25°, 35° and 40° are grouped with normal AOI (theta 0°) and quartz glass reference and shown in figure 2 (a) together, besides more AOIs among them (theta 5°, 10°, 20° and 30°) are displayed in the enlarged view of the trough bottom area in the inset. Although switching of AOI occurs, main features of filter spectrum for individual model are almost at same wavelength compared to that for normal AOI (theta 0° model). In the inset, the lowest of curves are at 532 nm wavelength except theta 30° curve which shift 1 nm to the right. Considering of absolute value in trough bottom area, it is found that transmission difference among these models paired with multiple AOIs is extremely small at 532 nm wavelength, e.g. only 0.03 between theta 0° and theta 30° models. If simply setting of low transmission at 532 nm wavelength as criterion, best blocking effect is generated at 35° AOI (theta 35°) whose transmission is 0.03 at 532 nm wavelength. Meanwhile, a trough of 40° AOI curve (black dash line) is also seen at 532 nm wavelength in figure 2 (a) and its inset, however another trough existing on its right and much lower than it. For this reason, we think that metamaterial filter for blocking of 532 nm light has been failed at this AOI, and angular tolerance of current device are ±35°.

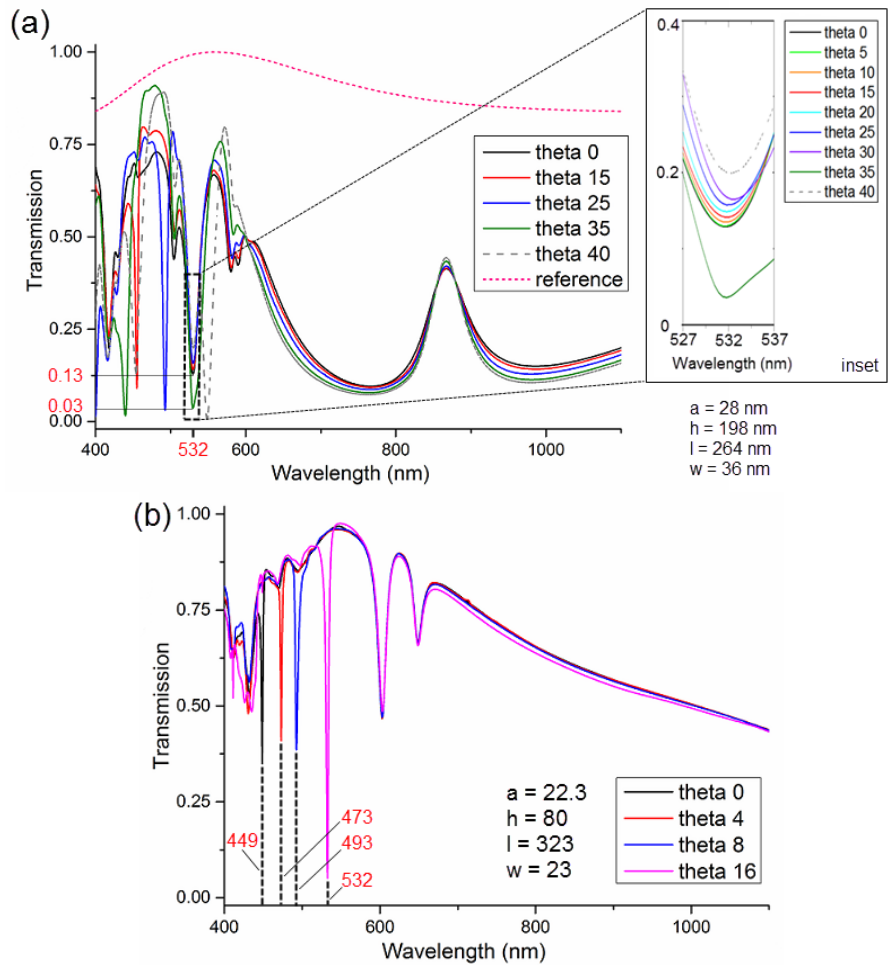

Figure 2 (a) Transmission curves of metamaterial filter with geometric parameters, a = 28 nm, h = 198 nm, l = 264 nm and w = 36 nm at multiple AOIs (theta) and pure quartz glass model as reference. Inset: Enlarged view of green light 532 nm wavelength area (b) Transmission curves of same shape model with geometric parameters, a = 22.3, h = 80, l = 323 and w =23 at four different AOIs

### 4. Discussion
#### 4.1 Wide-angle capability

It is known that traditional narrow-band-stop filters made of thin coatings/films does not possess wide-angle capability, and blue shift of spectrum features would occur with the growth of AOI [16, 17]. However, after comparison with numerous combinations of geometric parameters of Ag cross, we find that wide-angle capability will only be given to the metamaterial with particular sized unit cell defining by cross length, *l*, which is proportional to the targeted blocking wavelength. For this reason, apart from combination of geometric parameters mentioned in figure 2 (a), some others also have similar band-stop effect at 532 nm wavelength although there is lacking of wide-angle performance. Figure 2 (b) indicates that transmission curves of same cross shaped metamaterial with parameter combination a = 22.3 nm, h = 80 nm, l = 323 nm and w = 23 nm at four different AOIs, theta = 0 °, 4 °, 8 ° and 16 °. It is shown that curve features over 550 nm wavelength can be maximum maintained at four AOIs, however obvious red shift happens for the trough below 550 nm wavelength with the increase of AOI. Wavelengths of troughs are 449 nm, 473 nm, 493 nm and 532 nm respectively for theta 0 °, 4 °, 8 ° and 16 °. As a result, after multi-times calculation, it is known that wide-angle capability can be observed when $l$ is smaller than 300 nm, and ratio between blocking targeted wavelength and $l$ is about 1.8 - 2.0 depending on features on demand. Contrarily, if unit cell is large leading to small above mentioned ratio, wide-angle capability will be disappeared, instead of red shift with the increase of AOI as shown in figure 2 (b), which is different from the blue shift of traditional film/coating filter. At non-normal AOIs, light has to travel longer distance in the glass substrate compared with that at normal AOI, which results in various refractive indexes of glass, and interference mechanism working on traditional film/coating filter [1] is impossible to cancel out this discrepancy. Furthermore, wide-angle capability of proposed metamaterial structures is due to that alike effect of collective oscillation of free electron in Ag cross can be excited at multiple AOIs, which leads that power flow of light circulates the Ag cross (up to thousands) above substrate instead of normal penetration of glass [18].

### 4.2 Power flow near field analysis

Above mentioned wide-angle capability is relating to the resonance status of electrons at particular light wavelength. Figure 3 shows the power flow diagram of cross section (*zx* plane, in the middle of Ag cross) of model used to plot figure 2 (a) in near field 532 nm and 558 nm at different AOIs. Distinct band filter effects of current model at 532 nm wavelength are respectively represented by the diagram of theta 0 °, 20 ° and 50 ° in Figure 3 (a-c). In figure 3 (a-b), it is found that streams with high electric field intensity (red colour) are circulating and surrounding Ag cross (pink line window), while relatively low intensity streams (green colour) outflow from the glass substrate (black line window). However, similar phenomenon does not appear in the theta 50 ° diagram shown in figure 3 (c), and it is seen that power flow passes whole structure along 50 ° AOI. It means that within angular tolerance of wide-angle capability (0-35 ° AOI), Ag cross structure is able to maintain power flow circulation most of time. A main stream of power flow can be observed at the interface between Ag cross and glass, which is more obvious in figure 3 (b). When AOI is out of tolerance, e.g. theta = 50 ° in figure 3 (c), this light circulation and corresponding plasmonics are out of balance, which causes high power streams cannot be formed at the interface. This trend towards to the circumstance that metamaterial filter is irradiated by 558 nm wavelength light shown in figure 3 (d). There is no power flow circulation observed in figure 3 (d), and its transmission is reported over 0.6 shown in figure 2 (a), which means band-stop capability is significantly related to power flow circulation illustrated in figure 3.

From figure 3 it can be noticed that there are several singular points (vortexes and edge with opposite directions) located around the corners of Ag cross and interface with glass. The phase trajectories in the vicinity of the singular points form the clockwise vortexes which are in the relatively low electric intensity areas. It is known that phase trajectories of the system contain one and a half degree of freedom in the vicinity of singular points, and related clockwise vortex shape represents a stable focus in the phase space [19]. Also, power flow couples to the other planes through these singular points [20], which results in relatively low field intensity in these areas (blue colour). These singular points, especially at the two ends of Ag cross, could be the reason to form the high power stream at interface between silver cross and glass. Extreme non-normal AOI will complicate distribution of vortexes in local area, such as figure 3 (c), or reduce their number such as figure 3 (d), then inhibit the formation of main stream of power flow at interface more or less.

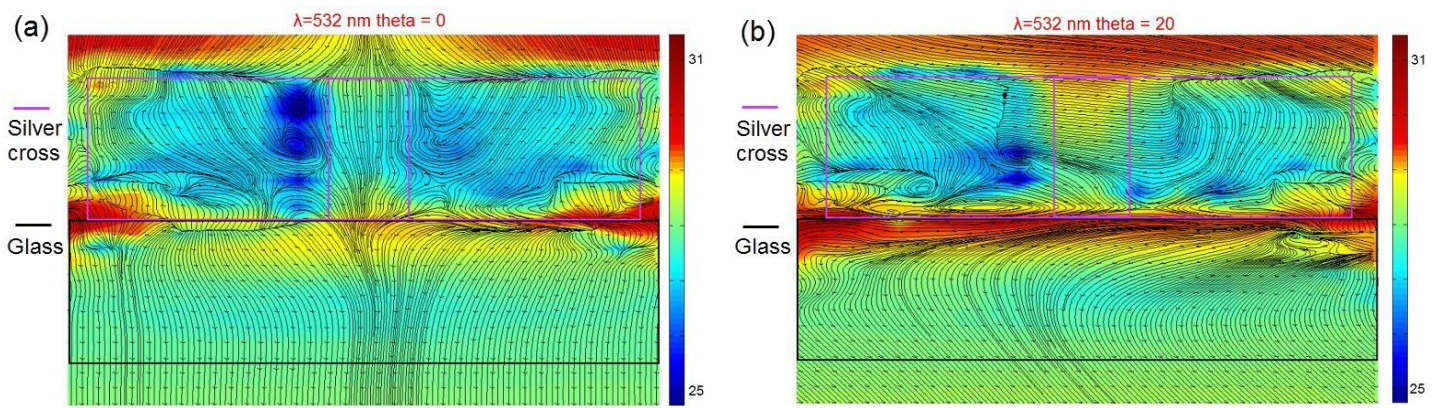

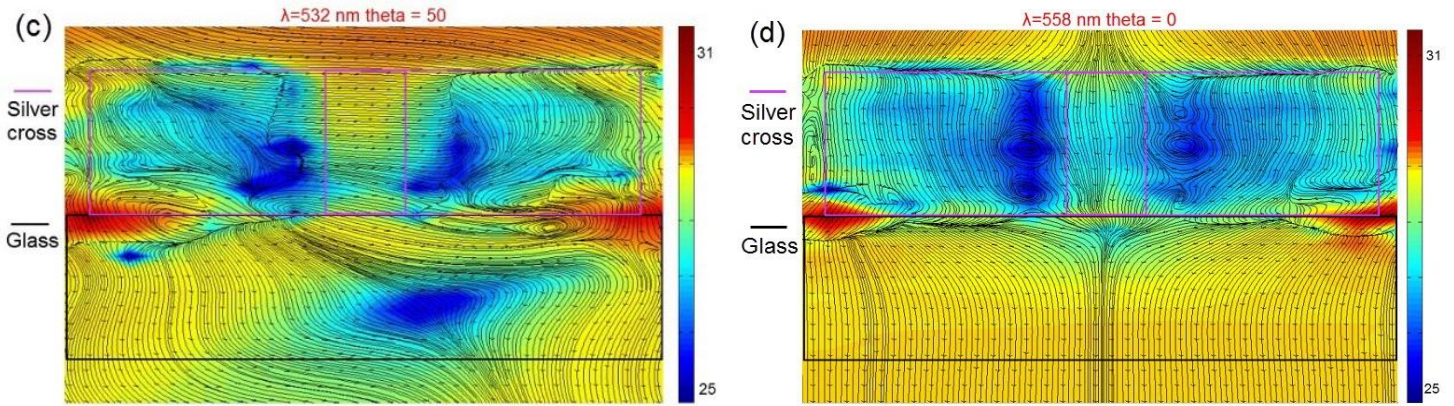

Figure 3 Power flow diagram for the model a = 28 nm, h = 198 nm, l = 264 nm and w = 36 nm at 532 nm wavelength (a) theta = 0 °(b) theta = 20 °(c) theta = 50 °and (d) at 558 nm wavelength theta = 0 °

## 5. Conclusion

In this paper, a cross shaped metamaterial is successfully simulated to perform as a wide-angle metamaterial narrow-band-stop filter. Low transmission (about 0.1 or less) can be provided by this model for 532 nm wavelength green light at ±35 °non-normal AOIs. After near field analysis it is found that cross shaped unit cell is able to circulate power flow at 532 nm wavelength to achieve band filter purpose, and regarding resonance can be maintained no matter AOI in the angular tolerance due to a major high power stream at the interface between Ag cross and glass caused by power flow vertexes locate at the Ag cross corners. This simulation is meaningful for development of a metamaterial device to carry out laser protection in aviation environment.

**Acknowledgement**

The authors gratefully acknowledge the financial support provided by the Sêr Cymru National Research Network in Advanced Engineering and Materials (ref: NRNF66 and NRN113) and Newton Research Collaboration Programme (ref: NCRP1516/1/153).